\documentclass[aps,prb,reprint,groupedaddress]{revtex4-2}
\usepackage{amsmath}
\usepackage{tikz}
\usepackage{graphicx}
\usepackage{dcolumn}
\usepackage{bm}
\usepackage{hyperref}
\usepackage{float}
\usepackage{soul}
\usepackage{ctex}

\begin{document}

\title{Transport Properties of a Quantum Dot Restudied by Algebraic Equation of Motion}

\author{Jiangqi Mao}
\author{Houmin Du}
\author{Yuliang Liu}
\email{ylliu@ruc.edu.cn}
\affiliation{Department of Physics, Renmin University of China, Beijing 100872, P.R.China}

\begin{abstract}
	Based on the algebraic equation of motion (AEOM) method, we investigate the transport properties of a quantum dot. We obtain an analytical expression for the dot electron single-particle Green's function, and based on this expression, we plot the dot electron density of states under different biases. We find that the Kondo resonance splits and is suppressed as the bias is increased. In addition, we calculate the differential conductance of the dot and obtain the zero-bias Kondo resonance at different temperatures, which is found to be suppressed as the temperature is increased.	\begin{description}
		\item[PACS]
		75.20.Hr,71.27.+a,03.65.Fd
	\end{description}
\end{abstract}

\maketitle

\section{Introduction}
The study of impurity quantum phase transition has been an active area in condensed matter physics. A impurity quantum phase transition, where only the degrees of freedom of subspace become critical, can be realized by tuning external parameters in the Hamiltonian rather than temperature\cite{vojta2006impurity}. Thus, the semiconductor quantum dot systems are a suitable platform to experimentally access different phases by identifying the transport bebaviors\cite{chang2009kondo}.

In terms of interacting Landauer formula\cite{meir1992landauer}, the transport of quantum dot systems is dependent on the density of states of local region. Due to the presence of strongly correlated electrons in quantum dots, the local density of states can be calculated by solving the Anderson model\cite{anderson1961localized} which effectively describes above systems. However, though the simplest case of a single quantum dot, correspondingly the single impurity Anderson model, has been well studied, solving it is a nontrivial thing\cite{hewson1997kondo, meir1993low, oguri2022current, Lacroix_1981, PhysRevB.59.9710, PhysRevB.52.10689, kashcheyevs2006applicability, PhysRevB.97.165140}.

The one usual way to solving the Anderson model are equation of motion approaches each of which there are different decoupling procedures\cite{Lacroix_1981,PhysRevB.59.9710,PhysRevB.52.10689,kashcheyevs2006applicability,PhysRevB.97.165140}. The common feature of these decouplings is that the high-order Green function is decomposed into the product of low-order Green function and the static correlation function which needs to be solved self-consistently in terms of fluctuation-dissipation theorem. In this way, the dot (single-particle) Green function is actually the solution of the integral equation. However, it is suspicious that this type of decoupling can be applied to strong correlated systems. On the other hand, with more higher-order Green functions appearing, the whole set of equations have no clear structure, which leads to the arbitrariness to some extent in the decoupling procedure.

Recently, we have proposed the algebraic equation of motion (EOM) approach\cite{doi:10.1142/S0217979218502582,doi:10.1142/S0217979219503557,doi:10.1142/S0217979221500648,Du_2023} to overcome difficulties mentioned above. There are two salient features in this approach: one is that we treat the high-order multiple-point correlation functions (see below) as a whole, which effectively incorporate the strong repulsive Coulomb interaction;the other is that all the AEOMs together constitute a hierarchical structure which can be classified by the parameter $N$ which labels the number of electrons residing in a length (sites) scale that take part in a time evolution of an electron from the initial to the final state. The multiple-point correlation functions in the same level $N$ constitute some linear equations about frequency in which there maybe a few multiple-point correlation functions belonging to the $N+1$ level. In order to make the whole AEOMs closed, we write the multiple-point correlation function to $N$ level, which capture the essence of physics, and then make cut-off approximation to the $N+1$ level appearing in $N$ level. Eventually, all we need to do is to perform an algebraic calculation on these linear AEOMs about frequency and analytically obtain the single-particle Green function. In this way, the static correlation functions will appear in the spectral weights without affecting the excitation spectrum, however, determining the corresponding values is beyond above approach.

In this paper our aim is to apply the AEOM approach to restudy the nonequilibrium transport in a quantum dot in terms of analytical expression of the dot Green function by solving the Anderson model. As a consequence, we find that the Kondo resonance in the dot density of states splits and is suppressed as the system deviates from equilibrium. Furthermore, we use the Landauer formula to calculate the differential conductance of the dot and obtain the zero-bias Kondo resonance, which is the primary signature of the Kondo effect in the quantum dot.

\section{method}
\subsection{formalism}

In order to explicitly show this approach, in this paper we apply AEOM to restudy the transport through a quantum dot in nonequilibrium. The single quantum dot system modeled by the Anderson Hamiltonian is as follows,
\begin{equation}\begin{aligned}
		H&=\sum_{\alpha ij\sigma}h_{\alpha ij}\hat{c}_{\alpha i\sigma}^{\dagger}\hat{c}_{\alpha j\sigma}	+\sum_{\sigma}\epsilon_{f}\hat{f}_{\sigma}^{\dagger}\hat{f}_{\sigma}+U\hat{n}_{f\uparrow}\hat{n}_{f\downarrow}\\
		&+\sum_{\alpha\sigma}(V_{\alpha}\hat{c}_{\alpha0\sigma}^{\dagger}\hat{f}_{\sigma}+V_{\alpha}^{*}\hat{f}_{\sigma}^{\dagger}\hat{c}_{\alpha0\sigma}).\label{eq1}
\end{aligned}\end{equation}
The first term on the right-hand side describes the electrons in left and right leads in a non-interacting conduction band, where $\hat{c}_{\alpha i\sigma}^{\dagger}$($\hat{c}_{\alpha i\sigma}$) creates(annihilates) an electron in conduction band with spin $\sigma=\uparrow,\downarrow$ at position $\mathbf{x}_{\alpha i}$ in each lead in which $\alpha=L,R$, $h_{\alpha ij}$ represents the nearest hopping which defined as $-t(\delta_{i,i\pm1}+\delta_{i\pm1,i})$. The second term describes the quantum dot, where $\epsilon_{f}$ is discrete energy level, and $U$ the on-site repulsive Coulomb interaction, and $\hat{f}_{\sigma}^{\dagger}(\hat{f}_{\sigma})$ the creation(annihilation) operator of the electrons on the quantum dot. The third term describes the tunneling between quantum dot and leads, and $V_{\alpha}$ is the tunneling amplitude, which assumed to be constant in this article.

Now we introduce a set of operators and calculate the corresponding commutators with above Hamiltonian. Interestingly, this set of commutation relations is closed to some extent, that is to say, no new individual operator appears. Firstly, the first set of commutation relations is as follows(we have used the Einstein summation convention),
\begin{equation}
	[\hat{n}_{f\bar{\sigma}},\hat{H}]=V_{\alpha}\hat{X}_{\alpha0\bar{\sigma}}^{(-)},
\end{equation}
\begin{equation}
	[\hat{c}_{\alpha i\sigma},\hat{H}]=h_{\alpha ij}\hat{c}_{\alpha j\sigma}+V_{\alpha}\delta_{i0}\hat{f}_{\sigma},
\end{equation}
\begin{equation}
	[\hat{f}_{\sigma},\hat{H}]=\epsilon_{f}\hat{f}_{\sigma}+U\hat{n}_{f\bar{\sigma}}\hat{f}_{\sigma}+V_{\alpha}\hat{c}_{\alpha0\sigma}
\end{equation}
where we have defined the composite operators $\hat{X}_{\alpha i\bar{\sigma}}^{(\pm)}=\hat{f}_{\bar{\sigma}}^{\dagger}\hat{c}_{\alpha i\bar{\sigma}}\pm\hat{c}_{\alpha i\bar{\sigma}}^{\dagger}\hat{f}_{\bar{\sigma}}$. In this spirit, we define in addition another composite operators $\hat{Y}_{\beta j\alpha i\bar{\sigma}}^{(\pm)}=\hat{c}_{\beta j\bar{\sigma}}^{\dagger}\hat{c}_{\alpha i\bar{\sigma}}\pm\hat{c}_{\alpha i\bar{\sigma}}^{\dagger}\hat{c}_{\beta j\bar{\sigma}}$. Then we explicitly write out the commutation relations about composite operators,
\begin{equation}\begin{aligned}
		[\hat{Y}_{\beta j\alpha i\bar{\sigma}}^{(\pm)},H]&=-h_{\beta jm}\hat{Y}_{\beta m\alpha i\bar{\sigma}}^{(\mp)}-V_{\beta}\delta_{j0}\hat{X}_{\alpha i\bar{\sigma}}^{(\mp)}\\
		&+h_{\alpha im}\hat{Y}_{\beta j\alpha m\bar{\sigma}}^{(\mp)}\mp V_{\alpha}\delta_{i0}\hat{X}_{\beta i\bar{\sigma}}^{(\mp)},
\end{aligned}\end{equation}
\begin{equation}\begin{aligned}
		[\hat{X}_{\alpha i\bar{\sigma}}^{(+)},\hat{H}]&=-(\epsilon_{f}+U\hat{n}_{f\sigma})\hat{X}_{\alpha i\bar{\sigma}}^{(-)}+h_{\alpha ij}\hat{X}_{\alpha j\bar{\sigma}}^{(-)}\\
		&-V_{\beta}\hat{Y}_{\beta 0\alpha i\bar{\sigma}}^{(-)},
\end{aligned}\end{equation}
\begin{equation}\begin{aligned}
		[\hat{X}_{\alpha i\bar{\sigma}}^{(-)},\hat{H}]&=-(\epsilon_{f}+U\hat{n}_{f\sigma})\hat{X}_{\alpha i\bar{\sigma}}^{(+)}+h_{\alpha ij}\hat{X}_{\alpha j\bar{\sigma}}^{(+)}\\
		&-V_{\beta}\hat{Y}_{\beta 0\alpha i\bar{\sigma}}^{(+)}+2\delta_{i0}V_{\alpha}\hat{n}_{f\bar{\sigma}}.
\end{aligned}\end{equation}
Indeed, the above set of commutation relations are closed to some extent.

The equations of motion are constructed by differentiating single-particle Green function in Heisenberg picture with time, the impurity Green function is defined as that
\begin{equation}
	G_{f\sigma}(t_1,t_2)=-i\langle\hat{T}\hat{f}_{\sigma}(t_1)\hat{f}_{\sigma}^\dagger(t_2)\rangle
\end{equation}
where $\langle\cdots\rangle$ represents the average with ground state. In terms of Heisenberg equation of motion of operators, we can utilize above a set of closed commutation relations to define the multiple-point correlation function in any order, which can be formally written as follows 
\begin{equation}\begin{aligned}
		F_{\{a\}mq}^{(\{A\})}(t_{1},t_{2})=-i\left<\hat{T}\prod_{k=1}^{N}\left[\hat{A}_{a}(t_{1})\right]^{k}\hat{f}_{m\sigma}(t_{1})\hat{f}_{q\sigma}(t_{2})\right>,\\
		\tilde{F}_{\{a\}mq}^{(\{A\})}(t_{1},t_{2})=-i\left<\hat{T}\prod_{k=1}^{N}\left[\hat{A}_{a}(t_{1})\right]^{k}\hat{c}_{m\sigma}(t_{1})\hat{f}_{q\sigma}(t_{2})\right>
\end{aligned}\end{equation}
where $\hat{A}=\{\hat{n},\hat{X}^{(\pm)},\hat{Y}^{(\pm)}\}$, and $a$ denotes the lattice and spin index for each operator $\hat{A}_{a}$, and $N$ is the number of operators appearing in the correlation functions. Based on these definitions, we can construct a hierarchical structure of algebraic equations labeled by the parameter $N$ in frequency domain, where the equations belonging to the same $N$ level form a subset of the entire equations, and in some cases there will be one or more $N+1$ level multiple-point correlation functions appearing in this subset but without $N-1$ level.

With the help of above set of commutation relations and the definition of multiple-point correlation function, we can directly write out the AEOM of corresponding multiple-point correlation function in any level.
\subsection{applying to single quantum dot}
In order to explicitly show the hierarchical structure of AEOMs, we write out AEOMs in $N=1$ level (here after we work out in frequency domain),
\begin{equation}
	\omega G_{\sigma}(\omega)=1+\epsilon_{f}G_{\sigma}(\omega)+UF_{\sigma}^{(n_{\bar{f}})}(\omega)+V_{\alpha}\tilde{G}_{\alpha0\sigma}(\omega),\label{eq10}
\end{equation}
\begin{equation}
	\omega\widetilde{G}_{\alpha i\sigma}(\omega)=h_{\alpha ij}\widetilde{G}_{\alpha j\sigma}(\omega)+V_{\alpha}\delta_{i0}G_{\sigma}(\omega),\label{eq11}
\end{equation}
\begin{equation}
	\begin{aligned}
		\omega F_{\sigma}^{(n_{\bar{f}})}(\omega)  &=  \left<\hat{n}_{f\bar{\sigma}}\right>+(\epsilon_{f}+U)F_{\sigma}^{(n_{\bar{f}})}(\omega)\\
		& +V_{\alpha}F_{\alpha0\sigma}^{(X_{\bar{\sigma}}(-))}(\omega)+V_{\alpha}\widetilde{F}_{\alpha0\sigma}^{(n_{\bar{f}})}(\omega),\label{eq12}
	\end{aligned}
\end{equation}
\begin{equation}
	\begin{aligned}
		\omega\widetilde{F}_{\alpha i\sigma}^{(n_{\bar{f}})}(\omega)  &=  V_{\beta}\widetilde{F}_{\beta0\alpha i\sigma}^{(X_{\bar{\sigma}}(-))}(\omega)+h_{\alpha ij}\widetilde{F}_{\alpha j\sigma}^{(n_{\bar{f}})}(\omega)\\
		& +V_{\alpha}\delta_{i0}F_{\sigma}^{(n_{\bar{f}})}(\omega),\label{eq13}
	\end{aligned}
\end{equation}
\begin{equation}
	\begin{aligned}
\omega F_{\alpha i\sigma}^{(X_{\bar{\sigma}}(-))}(\omega)  &=  -\epsilon_{f}F_{\alpha i\sigma}^{(X_{\bar{\sigma}}(+))}(\omega)-UF_{\alpha i\sigma}^{(n_{f}X_{\bar{\sigma}}(+))}(\omega)\\
&+h_{\alpha ij}F_{\alpha j\sigma}^{(X_{\bar{\sigma}}(+))}(\omega)-V_{\beta}F_{\beta\alpha0i\sigma}^{(Y_{\bar{\sigma}}(+))}(\omega)\\
    &+2\delta_{i0}V_{\alpha}F_{\sigma}^{(n_{\bar{f}})}(\omega)+\epsilon_{f}F_{\alpha i\sigma}^{(X_{\bar{\sigma}}(-))}(\omega)\\
   & +UF_{\alpha i\sigma}^{(X_{\bar{\sigma}}(-)n_{\bar{f}})}(\omega)+V_{\beta}\widetilde{F}_{\alpha i\beta0\sigma}^{(X_{\bar{\sigma}}(-))}(\omega)\label{eq14}
	\end{aligned}
\end{equation}
where $\widetilde{G}_{\alpha i\sigma}(t_1,t_2)=-i\langle\hat{T}\hat{c}_{\alpha i\sigma}(t_1)\hat{f}_{\sigma}^\dagger(t_2)\rangle$, $F_{\sigma}^{(n_{\bar{f}})}(t_1,t_2)=-i\langle\hat{T}\hat{n}_{f\bar{\sigma}}(t_1)\hat{f}_{\sigma}(t_1)\hat{f}_{\sigma}^{\dagger}(t_2)\rangle$, $\widetilde{F}_{\alpha i\sigma}^{(n_{\bar{f}})}(t_1,t_2)=-i\langle\hat{T}\hat{n}_{f\bar{\sigma}}(t_1)\hat{c}_{\alpha i\sigma}(t_1)\hat{f}_{\sigma}^{\dagger}(t_2)\rangle$ and $F_{\alpha i\sigma}^{(X_{\bar{\sigma}}(-))}(t_1,t_2)=-i\langle\hat{T}\hat{X}_{\alpha i\bar{\sigma}}^{(-)}(t_1)\hat{f}_{\sigma}(t_1)\hat{f}_{\sigma}^{\dagger}(t_2)\rangle$ are the above corresponding multiple-point correlation function in time domain. Up to now, if we make cut-off approximation by discarding the $F_{\alpha i\sigma}^{(X_{\bar{\sigma}}(-))}(\omega)$, then by a simple algebraic calculation, we have a Hubbard-I solution which corresponds the physics of Coulomb blockage regime,
\begin{equation}\begin{aligned}
G_{\sigma}(\omega)&=\frac{1-\left<\hat{n}_{f\bar{\sigma}}\right>}{\omega-\epsilon_{f}-\sum_{\alpha}V_{\alpha}^{2}\Gamma_{\alpha00}^{(-)}(\omega)}\\
&+\frac{\left<\hat{n}_{f\bar{\sigma}}\right>}{\omega-\epsilon_{f}-U-\sum_{\alpha}V_{\alpha}^{2}\Gamma_{\alpha00}^{(-)}(\omega)},
\label{HF}
\end{aligned}\end{equation}
where $\Gamma_{\alpha00}^{(\pm)}(\omega)=\frac{1}{N}\sum_{k}\frac{1}{\omega\pm\epsilon_{\alpha k}}$.

In order to figure out the Kondo physics, we must go further to incorporate the spin-flip Green function $\widetilde{F}_{\beta j\alpha i\sigma}^{(X_{\bar{\sigma}}(-))}(\omega)$ which appears in Eq.\eqref{eq14}, the corresponding AEOMs are as follows,
\begin{equation}
	\begin{aligned}
\omega\widetilde{F}_{\beta j\alpha i\sigma}^{(X_{\bar{\sigma}}(-))}(\omega) &=  -\epsilon_{f}\widetilde{F}_{\beta j\alpha i\sigma}^{(X_{\bar{\sigma}}(+))}(\omega)-U\widetilde{F}_{\beta j\alpha i\sigma}^{(n_{f}X_{\bar{\sigma}}(+))}(\omega)\\
&+h_{\beta jm}\widetilde{F}_{\beta m\alpha i\sigma}^{(X_{\bar{\sigma}}(+))}(\omega) -V_{\gamma}\widetilde{F}_{\gamma\beta0j\alpha i\sigma}^{(Y_{\bar{\sigma}}(+))}(\omega)\\
&+2\delta_{j0}V_{\beta}\widetilde{F}_{\alpha i\sigma}^{(n_{\bar{f}})}(\omega)+h_{\alpha im}F_{\beta j\alpha m\sigma}^{(X_{\bar{\sigma}}(-))}(\omega)\\
   & +V_{\alpha}\delta_{i0}F_{\beta j\sigma}^{(X_{\bar{\sigma}}(-))}(\omega),
   \label{FX-}
	\end{aligned}
\end{equation}
\begin{equation}
	\begin{aligned}
\omega\widetilde{F}_{\beta j\alpha i\sigma}^{(X_{\bar{\sigma}}(+))}(\omega) &=  -\epsilon_{f}\widetilde{F}_{\beta j\alpha i\sigma}^{(X_{\bar{\sigma}}(-))}(\omega)-U\widetilde{F}_{\beta j\alpha i\sigma}^{(n_{f}X_{\bar{\sigma}}(-))}(\omega)\\
&+h_{\beta jm}\widetilde{F}_{\beta m\alpha i\sigma}^{(X_{\bar{\sigma}}(-))}(\omega) -V_{\gamma}\widetilde{F}_{\gamma\beta0j\alpha i\sigma}^{(Y_{\bar{\sigma}}(+))}(\omega)\\
&+h_{\alpha im}F_{\beta j\alpha m\sigma}^{(X_{\bar{\sigma}}(+))}(\omega) +V_{\alpha}\delta_{i0}F_{\beta j\sigma}^{(X_{\bar{\sigma}}(+))}(\omega).\label{FX+}
	\end{aligned}
\end{equation}
where $\widetilde{F}_{\beta j\alpha i\sigma}^{(X_{\bar{\sigma}}(\pm))}(t_1,t_2)=-i\langle\hat{T}\hat{X}_{\beta j\bar{\sigma}}^{(\pm)}(t_1)\hat{c}_{\alpha i\sigma}(t_1)\hat{f}_{\sigma}^{\dagger}(t_2)\rangle$. We are not going to write out all the equations of motion of multiple point correlation functions appearing in above equations and put them in Appendix. By incorporating the spin-flip process which is a many-body effect between electrons in local region and leads, we can obtain the physics of Kondo regime. It is worth noting that the higher-order spin-flip Green function $\widetilde{F}_{\beta j\alpha i\sigma}^{(n_{f}X_{\bar{\sigma}}(\pm))}(\omega)$ appearing in Eq.(\ref{FX-},\ref{FX+}) which contributes the corresponding Kondo resonance spectral weight, in low energy limit we have
\begin{equation}
	\widetilde{F}_{\beta j\alpha i\sigma}^{(n_{f}X_{\bar{\sigma}}(\mp))}(0)=\mp\frac{C_{\beta j\alpha i}}{\epsilon_{f}+U}\label{FnX}
\end{equation}
where $C_{\alpha i\beta j}=\frac{1}{2}\left<\left[\hat{S}^{(+)}\hat{s}_{\beta j,\alpha i}^{(-)}+\hat{S}^{(-)}\hat{s}_{\beta j,\alpha i}^{(+)}\right]\right>$, $\hat{s}_{\beta j,\alpha i}^{(+)}=\hat{c}_{\beta j\uparrow}^{\dagger}\hat{c}_{\alpha i\downarrow}$, $\hat{s}_{\beta j,\alpha i}^{(-)}=\hat{c}_{\beta j\downarrow}^{\dagger}\hat{c}_{\alpha i\uparrow}$, $\hat{S}^{(+)}=\hat{f}_{\downarrow}^{\dagger}\hat{f}_{\uparrow}$ and $\hat{S}^{(-)}=\hat{f}_{\uparrow}^{\dagger}\hat{f}_{\downarrow}$, which is static correlation function representing the spin-flip process between electrons in local region and leads. However, the quantitative calculation of $C_{\alpha i\beta j}$ is beyond the scope of current approach. In this paper, we treat it as a parameter.

\section{results}
The calculation procedure employed here is the same as what we did before\cite{Du_2023}, the analytical expression of the dot Green function is as follows, 
\begin{equation}\begin{aligned}
		G_{\sigma}(\omega)  &=  \frac{A}{\omega-\epsilon_{f}-V^{2}\Gamma_{\alpha00}^{(-)}(\omega)}
		+\frac{B}{\omega-[\epsilon_{f}+\eta(\omega)]-\Delta(\omega)}\\
		& +\frac{C}{\omega-\epsilon_{f}-U- V_{\alpha}^{2}\Gamma_{\alpha00}^{(-)}(\omega)}\label{G}
\end{aligned}\end{equation}
with 
\begin{equation}\begin{aligned}
		A&=  1-n_{f\bar{\sigma}}+\frac{Jn_{f\bar{\sigma}}}{2D(\omega)}\\
		& +\frac{1}{8}\frac{U^{2}\sum_{\alpha,\beta}C_{\alpha0\beta0}}{(U+\epsilon_{f})}\frac{J}{D(\omega)}[\frac{1}{ V_\alpha^{2}\Gamma_{\alpha00}^{(-)}(\omega)-\eta(\omega)-\Delta(\omega)}],
		\nonumber
\end{aligned}\end{equation}
\begin{equation}\begin{aligned}
	B  &= -\frac{1}{8}\frac{U^{2}\sum_{\alpha,\beta}C_{\alpha0\beta0}}{(U+\epsilon_{f})}\frac{J}{D(\omega)}\\&\times\frac{1}{U+ V_{\alpha}^{2}\Gamma_{\alpha00}^{(-)}(\omega)-(\Delta(\omega)+\eta(\omega))}\\
		& +\frac{1}{8}\frac{U^{2}\sum_{\alpha,\beta}C_{\alpha0\beta0}}{(U+\epsilon_{f})}\frac{J}{D(\omega)}\frac{1}{ V_{\alpha}^{2}\Gamma_{\alpha00}^{(-)}(\omega)-(\Delta(\omega)+\eta(\omega))},
		\nonumber
\end{aligned}\end{equation}
and 
\begin{equation}\begin{aligned}
		C  &= n_{f\bar{\sigma}}-\frac{Jn_{f\bar{\sigma}}}{D(\omega)}
		   +\frac{1}{8}\frac{U^{2}\sum_{\alpha,\beta}C_{\alpha0\beta0}}{(U+\epsilon_{f})}\frac{J}{D(\omega)}\\
		&\times\frac{1}{U+V_{\alpha}^{2}\Gamma_{\alpha00}^{(-)}(\omega)-(\Delta(\omega)+\eta(\omega))},
		\nonumber
\end{aligned}\end{equation}
where $n_{f\bar{\sigma}}=\left<{n}_{f\bar{\sigma}}\right>$ as a parameter, to determine its exact value is beyond our approach. The expression for $D(\omega)$ is given by $D(\omega)=[\omega-(\epsilon_{f}+\frac{U}{2})]+\frac{JU}{\epsilon_{f}+U+V_{\alpha}^{2}\Gamma_{\alpha00}^{(-)}(\omega)}$. We assume the coupling strength between the left and right leads and
the dot is the same, i.e.,$V_{L}=V_{R}=V$, and $J=\frac{4V^{2}}{U}$, which can be understood as the strength of the spin-spin coupling which arises naturally from the decoupling processes,$\Delta(\omega)=\frac{V_{\alpha}^{2}\Sigma_{\alpha00}^{(+)}(\omega)}{2}$, $\eta(\omega)=\frac{V_{\alpha}^{2}\Sigma_{\alpha00}^{(-)}(\omega)}{2}$, where $\Sigma_{\alpha00}^{(\pm)}(\omega)= \Lambda_{\alpha00}^{(-)}(\omega)\pm\Lambda_{\alpha00}^{(+)}(\omega) $ and $\Lambda_{\alpha00}^{(\pm)}(\omega)=\frac{1}{N}\sum_{k}\frac{\theta(\mu_\alpha-\epsilon_{\alpha k})}{\omega\pm\epsilon_{\alpha k}}$.

Observing Eq.\eqref{G} it is apparent that there are three poles, the first and last correspond to the physics of Coulomb blockage, the remaining the physics of Kondo regime, in terms of which we can deduce the logarithmic renormalized energy level and define the Kondo temperature, which same as the usual equation of motion approach\cite{Lacroix_1981}. The weight of Kondo resonance is incorporated in spin-flip correlation function $\sum_{\alpha,\beta}C_{\alpha0\beta0}$, which consists of four contributions: $C_{L0L0}$, $C_{L0R0}$, $C_{R0R0}$ and $C_{R0L0}$, which represents the four possible spin-flip processes between electrons in local region and leads. For simplicity, we only consider the average contribution and take them as a parameter $C_{00}=\sum_{\alpha,\beta}C_{\alpha0\beta0}$. 
\begin{figure}[H]
	\centering
	\includegraphics[scale=0.25]{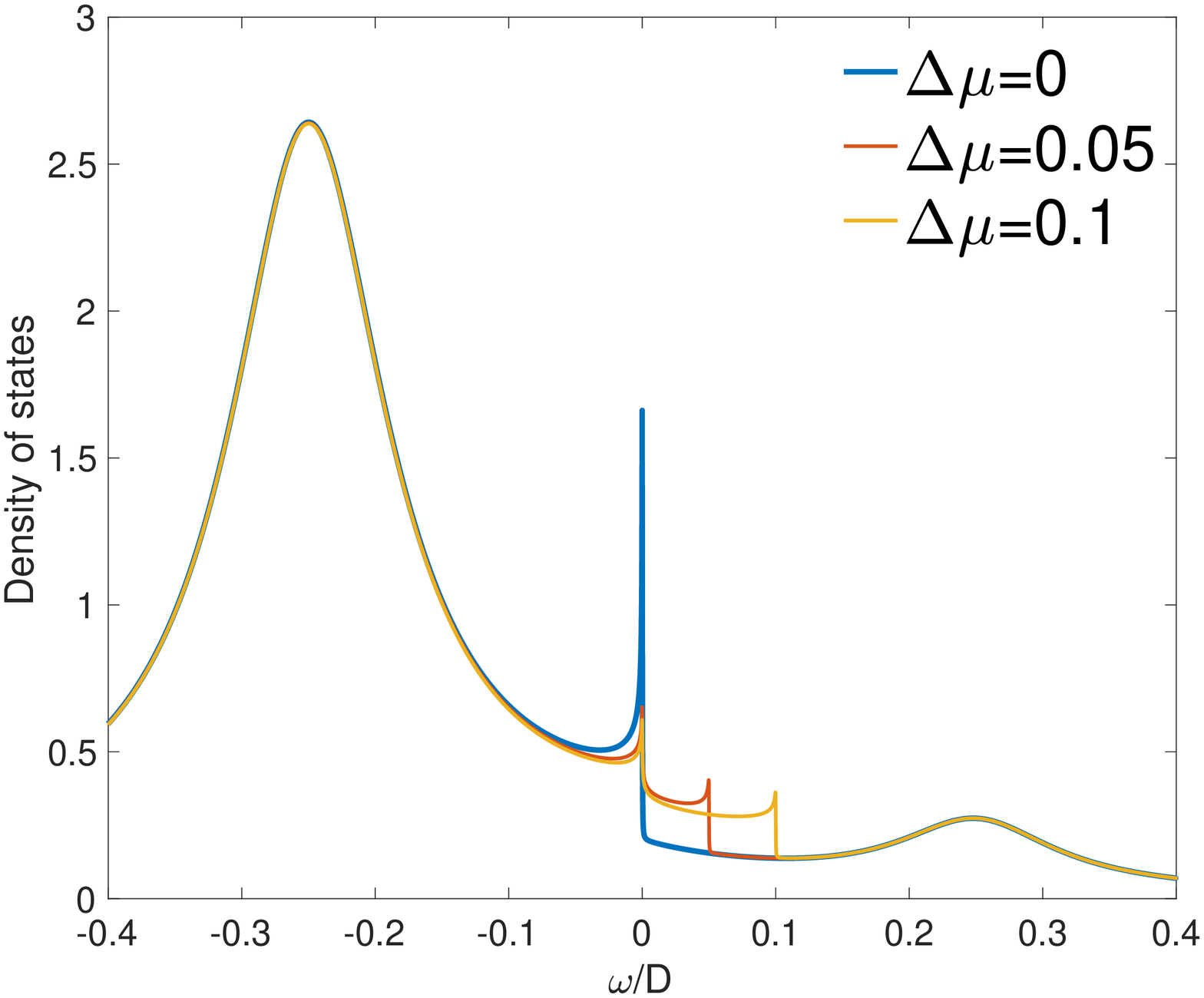}
	\caption{Density of states for local region on equilibrium and nonequilibrium. The corresponding parameters are $C_{00}=-0.4$, $V=0.15$ eV, $D=1$ eV, $U=0.5$ eV, $n_{f\bar{\sigma}}=0.5$ and $\rho=1/2D$. The Kondo temperature is close to 0. With the increase of the bias voltage eV$=\mu_L-\mu_R$, with $\mu_{R}=0$ fixed, the Kondo resonance splits and is suppressed.}
	\label{fig1}
\end{figure}

Following the Eq.\eqref{G} and considering the case of particle-hole symmetric case $\epsilon_f=-U/2$, in terms of $\rho_\sigma(\omega)=-\frac{1}{\pi}\text{Im}G_\sigma(\omega+i0^+)$ we obtain the dot electron density of states. As shown in Fig.(\ref{fig1}), the density of states under different biases $\Delta\mu=\mu_{L}-\mu_{R}$ is displayed. When the system is in equilibrium, i.e.$\mu_{L}=\mu_{R}=0$, the famous three-peak structure is shown, where the Kondo resonance near the Fermi level is a sharp and narrow peak, and the two broad peaks located at $\pm U/2$ correspond to the energy levels $\epsilon_{f}$ and $\epsilon_{f}+U$, respectively. When the system out of equilibrium, the Kondo resonance splits and is suppressed. The splitting increases with increasing bias voltage. This behavior is the same as the results of ref.\cite{meir1993low}.

\begin{figure}[H]
	\centering
	\includegraphics[scale=0.27]{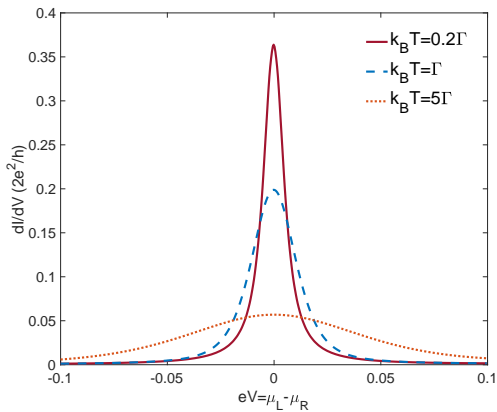}
	\caption{Differential conductance spectra of the quantum dot for different temperatures. A peak appears at zero bias voltage. The $\Gamma_{}=2\pi\rho V^2$ represents the coupling strength between the quantum dot and the leads. The parameter values are the same as in Figure (\ref{fig1}).}
	\label{fig2}
\end{figure}

According to interacting Landauer formula\cite{meir1992landauer}, the current through a quantum dot can be calculated as follows
\begin{equation}
J=\frac{e}{\hbar}\sum_{\sigma}\int_{-\infty}^{\infty}d\epsilon\frac{\Gamma_{L}\Gamma_{R}}{\Gamma_{L}+\Gamma_{R}}[f_{L}(\epsilon)-f_{R}(\epsilon)]\rho_{\sigma}(\epsilon)
\label{J}
\end{equation}
where $\Gamma_{\alpha}$ represents the coupling strength between the quantum dot and the $\alpha$ lead, and $f_{\alpha}(\epsilon)$ is the Fermi distribution function of electrons in leads, and $\rho_{\sigma}(\epsilon)$ is the density of states for the dot. In this paper, we consider the case where the quantum dot is symmetrically coupled to both leads, so $\Gamma_{\alpha}=2\pi\rho V^{2}$. We use this formula to calculate the current through quantum dot and obtain the differential conductance spectrum in Fig.(\ref{fig2}), which is same as the result by usual EOM approach\cite{meir1993low}.

\section{conclusions}
We have applied AEOM to restudy the transport properties of a quantum dot and analytically obtain the dot Green function which incorporates the low-lying excitation, that is the Kondo resonance. Different from the usual EOM approach in which there appears the self-consistent integral equations, in our frame, as decoupling the set of equations we only have to do is algebraic calculation. Though, the values of static correlations beyond present approach, they do not affect the excitation spectrum. We have found that the Kondo resonance splits and the splitting proportionates to the bias, meanwhile the corresponding weights are suppressed. In terms of interacting Landauer formula, we have also obtained the differential conductance spectra which is same as the previous result. Hence we conclude that our new approach will be suitable for handling the corresponding problems of double quantum dot systems.

\begin{acknowledgments}
	This work is supported by NSFC (Grant No.11974420).
\end{acknowledgments}

\appendix

\section{multiple-point correlation functions}

For Eq.\eqref{eq14}, in the Kondo limit, we discard the multiple-point correlation functions proportional to $V$ and consider the low-energy limit, which ultimately simplifies to
\begin{equation}\begin{aligned}
\omega F_{\alpha i\sigma}^{(X_{\bar{\sigma}}(-))}(\omega)=
V_{\beta}\widetilde{F}_{\alpha i\beta0\sigma}^{(X_{\bar{\sigma}}(-))}(\omega)+2\delta_{i0}V_{\alpha}F_{\sigma}^{(n_{\bar{f}})}(\omega)\label{A1}
\end{aligned}\end{equation}
\begin{widetext}
Similarly, by simplifying equations \eqref{FX-} and \eqref{FX+}, we obtain
\begin{equation}\begin{aligned}
\hbar\omega\widetilde{F}_{\beta j\alpha i\sigma}^{(X_{\bar{\sigma}}(-))}(\omega) &=  -\epsilon_{f}\widetilde{F}_{\beta j\alpha i\sigma}^{(X_{\bar{\sigma}}(+))}(\omega)-U\widetilde{F}_{\beta j\alpha i\sigma}^{(n_{f}X_{\bar{\sigma}}(+))}(\omega)+h_{\beta jm}\widetilde{F}_{\beta m\alpha i\sigma}^{(X_{\bar{\sigma}}(+))}(\omega)\\
  & -V_{\gamma}\widetilde{F}_{\gamma0\beta j\alpha i\sigma}^{(Y_{\bar{\sigma}}(+))}(\omega)+h_{\alpha im}\widetilde{F}_{\beta j\alpha m\sigma}^{(X_{\bar{\sigma}}(-))}(\omega),\label{A2}
\end{aligned}\end{equation}
\begin{equation}\begin{aligned}
\hbar\omega\widetilde{F}_{\beta j\alpha i\sigma}^{(X_{\bar{\sigma}}(+))}(\omega) &=  -\epsilon_{f}\widetilde{F}_{\beta j\alpha i\sigma}^{(X_{\bar{\sigma}}(-))}(\omega)-U\widetilde{F}_{\beta j\alpha i\sigma}^{(n_{f}X_{\bar{\sigma}}(-))}(\omega)+h_{\beta jm}\widetilde{F}_{\beta m\alpha i\sigma}^{(X_{\bar{\sigma}}(-))}(\omega)\\
  & -V_{\gamma}\widetilde{F}_{\gamma0\beta j\alpha i\sigma}^{(Y_{\bar{\sigma}}(-))}(\omega)+h_{\alpha im}\widetilde{F}_{\beta j\alpha m\sigma}^{(X_{\bar{\sigma}}(+))}(\omega).\label{A3}
\end{aligned}\end{equation}
Then we calculate the AEOM of $\widetilde{F}_{\gamma l\beta j\alpha i\sigma}^{(Y_{\bar{\sigma}}(\pm))}(\omega)$, which is given by
\begin{equation}\begin{aligned}
\omega\widetilde{F}_{\gamma l\beta j\alpha i\sigma}^{(Y_{\bar{\sigma}}(-))}(\omega)  = & -h_{\gamma lm}\widetilde{F}_{\gamma m\beta j\alpha i\sigma}^{(Y_{\bar{\sigma}}(+))}(\omega)-V_{\gamma}\delta_{l0}\widetilde{F}_{\beta j\alpha i\sigma}^{(X_{\bar{\sigma}}(+))}(\omega) +h_{\beta jm}\widetilde{F}_{\gamma l\beta m\alpha i\sigma}^{(Y_{\bar{\sigma}}(+))}(\omega)+V_{\beta}\delta_{j0}\widetilde{F}_{\gamma l\alpha i\sigma}^{(X_{\bar{\sigma}}(+))}(\omega)\\
   & +h_{\alpha im}\widetilde{F}_{\gamma l\beta j\alpha m\sigma}^{(Y_{\bar{\sigma}}(-))}(\omega)+V_{\alpha}\delta_{i0}F_{\gamma l\beta j\sigma}^{(Y_{\bar{\sigma}}(-))}(\omega),\label{A4}
\end{aligned}\end{equation}
\begin{equation}\begin{aligned}
\omega\widetilde{F}_{\gamma l\beta j\alpha i\sigma}^{(Y_{\bar{\sigma}}(+))}(\omega)  = & -h_{\gamma lm}\widetilde{F}_{\gamma m\beta j\alpha i\sigma}^{(Y_{\bar{\sigma}}(-))}(\omega)-V_{\gamma}\delta_{l0}\widetilde{F}_{\beta j\alpha i\sigma}^{(X_{\bar{\sigma}}(-))}(\omega)+h_{\beta jm}\widetilde{F}_{\gamma l\beta m\alpha i\sigma}^{(Y_{\bar{\sigma}}(-))}(\omega)+V_{\beta}\delta_{j0}\widetilde{F}_{\gamma l\alpha i\sigma}^{(X_{\bar{\sigma}}(-))}(\omega)\\
  & +h_{\alpha im}\widetilde{F}_{\gamma l\beta j\alpha m\sigma}^{(Y_{\bar{\sigma}}(+))}(\omega)+V_{\alpha}\delta_{i0}F_{\gamma l\beta j\sigma}^{(Y_{\bar{\sigma}}(+))}(\omega).\label{A5}
\end{aligned}\end{equation}
$\widetilde{F}_{\gamma l\beta j\alpha i\sigma}^{(Y{\bar{\sigma}}(\pm))}$ includes correlations between the three lead sites and the dot electron, namely $\gamma l$, $\beta j$, and $\alpha i$. By comparing with Eq.\eqref{A2} and Eq.\eqref{A3}, we only keep the multiple-point correlation functions $\widetilde{F}_{\gamma m\beta j\alpha i\sigma}^{(Y{\bar{\sigma}}(\pm))}(\omega)  $, $\widetilde{F}_{\gamma l\beta m\alpha i\sigma}^{(Y_{\bar{\sigma}}(\pm))}(\omega)$ and $\widetilde{F}_{\beta j\alpha i\sigma}^{(X{\bar{\sigma}}(\pm))}(\omega)$, then discard the others. Hence, we obtain the following equations,
\begin{equation}\begin{aligned}
\omega\widetilde{F}_{\gamma l\beta j\alpha i\sigma}^{(Y_{\bar{\sigma}}(-))}(\omega)+h_{\gamma lm}\widetilde{F}_{\gamma m\beta j\alpha i\sigma}^{(Y_{\bar{\sigma}}(+))}(\omega)-h_{\beta jm}\widetilde{F}_{\gamma k\beta m\alpha i\sigma}^{(Y_{\bar{\sigma}}(+))}(\omega)=-\delta_{l0}V_{\gamma}\widetilde{F}_{\beta j\alpha i\sigma}^{(X_{\bar{\sigma}}(+))}(\omega)\\
\omega\widetilde{F}_{\gamma l\beta j\alpha i\sigma}^{(Y_{\bar{\sigma}}(+))}(\omega)+h_{\gamma lm}\widetilde{F}_{\gamma m\beta j\alpha i\sigma}^{(Y_{\bar{\sigma}}(-))}(\omega)-h_{\beta jm}\widetilde{F}_{\gamma k\beta m\alpha i\sigma}^{(Y_{\bar{\sigma}}(-))}(\omega)=-\delta_{l0}V_{\gamma}\widetilde{F}_{\beta j\alpha i\sigma}^{(X_{\bar{\sigma}}(-))}(\omega)
\end{aligned}\end{equation}
In order to decouple above equations, we need to take twice Fourier transformations respect to each hopping term and then integrate out the conduction band. However, in doing so, $\widetilde{F}_{\beta j\alpha i\sigma}^{(X_{\bar{\sigma}}(\pm))}(\omega)$ can not be decoupled. To this end, we preserve the hole excitation and assume the particle excitation near the Fermi surface is zero, thus $\widetilde{F}_{\beta j\alpha i\sigma}^{(X_{\bar{\sigma}}(\pm))}(\omega)$ including particle excitation decouple, then substituting the result into Eq.\eqref{A2} and Eq.\eqref{A3}, we obtain
\begin{equation}\begin{aligned}
&\left[\omega-\Delta(\omega)\right] \widetilde{F}_{\beta j\alpha i\sigma}^{(X_{\bar{\sigma}}(-))}(\omega)+\left[\epsilon_{f}+\eta(\omega)\right]\widetilde{F}_{\beta j\alpha i\sigma}^{(X_{\bar{\sigma}}(+))}(\omega)=-U\widetilde{F}_{\beta j\alpha i\sigma}^{(n_{f}X_{\bar{\sigma}}(+))}(\omega),\label{A7}
\end{aligned}\end{equation}
\begin{equation}\begin{aligned}
&\left[\omega-\Delta(\omega)\right] \widetilde{F}_{\beta j\alpha i\sigma}^{(X_{\bar{\sigma}}(+))}(\omega)+\left[\epsilon_{f}+\eta(\omega)\right]\widetilde{F}_{\beta j\alpha i\sigma}^{(X_{\bar{\sigma}}(-))}(\omega)=-U\widetilde{F}_{\beta j\alpha i\sigma}^{(n_{f}X_{\bar{\sigma}}(-))}(\omega).\label{A8}
\end{aligned}\end{equation}
Here,$\Delta(\omega)=\frac{V_{\alpha}^{2}\Sigma_{\alpha00}^{(+)}(\omega)}{2}$, $\eta(\omega)=\frac{V_{\alpha}^{2}\Sigma_{\alpha00}^{(-)}(\omega)}{2}$, where $\Sigma_{\alpha00}^{(\pm)}(\omega)=\theta(-k)\left[  \Lambda_{\alpha00}^{(-)}(\omega)\pm\Lambda_{\alpha00}^{(+)}(\omega) \right] $ and $\Lambda_{\alpha00}^{(\pm)}(\omega)=\frac{1}{N}\sum_{k}\frac{1}{\omega\pm\epsilon_{\alpha k}}$.

The AEOMs of $\widetilde{F}_{\beta j\alpha i\sigma}^{(n{f}X_{\bar{\sigma}}(\pm))}(\omega)$ are as follows,
\begin{equation}\begin{aligned}
\omega\widetilde{F}_{\alpha i\beta j\sigma}^{(n_{f}X_{\bar{\sigma}}(-))}(\omega)&=\left<\{\hat{n}_{f\sigma}\hat{X}_{\alpha i\bar{\sigma}}^{(-)}\hat{c}_{\beta j\sigma},\hat{f}_{\sigma}^{\dagger}\}\right>
+V_{\gamma}\widetilde{F}_{\gamma0\alpha i\beta j\sigma}^{(X_{\sigma}(-)X_{\bar{\sigma}}(-))}(\omega)-\epsilon_{f}\widetilde{F}_{\alpha i\beta j\sigma}^{(n_{f}X_{\bar{\sigma}}(+))}(\omega)-U\widetilde{F}_{\alpha i\beta j\sigma}^{(n_{f}X_{\bar{\sigma}}(+))}(\omega)\\
&+h_{\alpha im}\widetilde{F}_{\alpha m\beta j\sigma}^{(n_{f}X_{\bar{\sigma}}(+))}(\omega)-V_{\gamma}\widetilde{F}_{\gamma0\alpha i\beta j\sigma}^{(n_{f}Y_{\bar{\sigma}}(+))}(\omega)+2\delta_{i0}V_{\alpha}\widetilde{F}_{\beta j\sigma}^{(n_{f})}(\omega)\\
&+h_{\beta jm}\widetilde{F}_{\alpha i\beta m\sigma}^{(n_{f}X_{\bar{\sigma}}(-))}(\omega)+V_{\beta}\delta_{j0}F_{\alpha i\sigma}^{(n_{f}X_{\bar{\sigma}}(-))}(\omega),\label{A9}
\end{aligned}\end{equation}
\begin{equation}\begin{aligned}
\omega\widetilde{F}_{\alpha i\beta j\sigma}^{(n_{f}X_{\bar{\sigma}}(+))}(\omega)&=\left<\{\hat{n}_{f\sigma}\hat{X}_{\alpha i\bar{\sigma}}^{(+)}\hat{c}_{\beta j\sigma},\hat{f}_{\sigma}^{\dagger}\}\right>
+V_{\gamma}\widetilde{F}_{\gamma0\alpha i\beta j\sigma}^{(X_{\sigma}(-)X_{\bar{\sigma}}(+))}(\omega)-\epsilon_{f}\widetilde{F}_{\alpha i\beta j\sigma}^{(n_{f}X_{\bar{\sigma}}(-))}(\omega)-U\widetilde{F}_{\alpha i\beta j\sigma}^{(n_{f}X_{\bar{\sigma}}(-))}(\omega)\\
&+h_{\alpha im}\widetilde{F}_{\alpha m\beta j\sigma}^{(n_{f}X_{\bar{\sigma}}(-))}(\omega)-V_{\gamma}\widetilde{F}_{\gamma0\alpha i\beta j\sigma}^{(n_{f}Y_{\bar{\sigma}}(-))}(\omega)\\
&+h_{\beta jm}\widetilde{F}_{\alpha i\beta m\sigma}^{(n_{f}X_{\bar{\sigma}}(+))}(\omega)+V_{\beta}\delta_{j0}F_{\alpha i\sigma}^{(n_{f}X_{\bar{\sigma}}(+))}(\omega).\label{A10}
\end{aligned}\end{equation}
\end{widetext}
Up to now, we have written out all the AEOMs in $N=1$ level and $N=2$ which capture the main physics. We do not go further and make a cut-off approximation to $N=2$ level. Then, we simply discard the terms that are not same as $\widetilde{F}_{\alpha i\beta j\sigma}^{(n_{f}X_{\bar{\sigma}}(\pm))}$ in Eq.\eqref{A9} and Eq.\eqref{A10}. Then, in the low-energy limit, we finally obtain the Eq.\eqref{FnX}. Substituting it back into Eq.\eqref{A7} and \eqref{A8} and combining with Eq.\eqref{eq10}-\eqref{eq13} and Eq.\eqref{A1}, we finally obtain the dot Green function.

\bibliographystyle{iopart-num.bst}
\bibliography{ref}

\end{document}